\documentclass[11pt,reqno]{amsart}
\usepackage{amsmath,amsgen,amstext,amsbsy,amsopn,amsthm,amssymb}

\newtheorem{thm}{THEOREM}

\theoremstyle{definition}
\newtheorem{rem}{REMARK}
\newtheorem{ex}{EXAMPLE}

\newcommand{\R}{{\mathbb R}}
\newcommand{\C}{{\mathbb C}}
\newcommand{\N}{{\mathbb N}}

\newcommand{\Hh}{{\mathcal H}}

\newcommand{\eps}{\varepsilon}

\newcommand{\x}{{x}}
\newcommand{\z}{{z}}
\newcommand{\y}{{y}}

\newcommand{\Tr}{{\rm Tr}}
\newcommand{\half}{\mbox{$\frac{1}{2}$}}

\newcommand{\dodd}{\delta_{\rm odd}}
\newcommand{\kF}{k_{\rm F}}
\newcommand{\const}{{\rm const. \,}}

\date{\small\version}

\begin{document}
\title[Decomposition of radial functions]
{General decomposition of radial\\ functions on $\R^n$ and 
applications to $N$-body quantum systems}
\author[Christian Hainzl]{Christian Hainzl$^1$}
\address{Mathematisches Institut, LMU M\"unchen, 
Theresienstrasse 39, 80333 Munich, Germany}
\email{hainzl@mathematik.uni-muenchen.de}
\author[Robert Seiringer]{Robert Seiringer$^2$}
\address{Department of Physics, Jadwin Hall, Princeton University,
P.O. Box 708, Princeton, New Jersey 08544, USA}
\email{rseiring@math.princeton.edu}
\date{\today}

\begin{abstract}
We present a generalization of the Fefferman-de la Llave decomposition
of the Coulomb potential to quite arbitrary radial functions $V$ on
$\R^n$ going to zero at infinity. This generalized decomposition can
be used to extend previous results on $N$-body quantum systems with
Coulomb interaction to a more general class of interactions. As an
example of such an application we derive the high density asymptotics
of the ground state energy of jellium with Yukawa interaction in the
thermodynamic limit, using a correlation estimate by Graf and Solovej
\cite{GS94}.
\end{abstract}

\maketitle

\footnotetext[1]{Marie Curie Fellow} 
\footnotetext[2]{Erwin Schr\"odinger Fellow. On leave from 
Institut f\"ur Theoretische Physik,
Universit\"at Wien, Boltzmanngasse 5, A-1090 Vienna,
Austria}

\section{Introduction}

For the description of physical systems interacting with Coulomb
forces, the Fefferman-de la Llave decomposition of the Coulomb
potential has pro\-ved very useful. It was introduced in \cite{FL86},
where it was used in the proof of stability of relativistic matter. It
states that, for $x\in\R^3$,
\begin{equation}\label{feff}
\frac 1{|x|}=\frac 1\pi \int_0^\infty dr \frac 1{r^5} \chi_r*\chi_r(x)\ ,
\end{equation}
where $\chi_r(x)=\theta(r-|x|)$ is the characteristic function of a ball
of radius $r$ centered at the origin, and $*$ denotes convolution on
$\R^3$. Except for the constant $1/\pi$, it is easily checked that
(\ref{feff}) holds true, since the right side is a radial,
homogeneous function of order $-1$. 

We are interested in a generalization of (\ref{feff}) to arbitrary
radial functions $V$ on $\R^n$ going to zero at infinity, with a
weight function $g(r)$ replacing $1/r^5$ in the integrand (see
(\ref{vg})). In Theorem~\ref{thm1} below we present a simple and
straightforward derivation of such a decomposition, assuming $V$ to
satisfy some decrease and regularity properties specified below. We
give an explicit expression for the weight function $g(r)$, which
turns out to be related to the $[n/2]+2$'th derivative of $V$. Here
$[\,\cdot\,]$ denotes the Gau\ss\ bracket, i.e., $[m]=\max\{n\in\N_0,
n\leq m\}$. In particular, the case of a positive $g$ is of interest,
implying positive definiteness of the function $V$.

Decompositions of the type (\ref{vg}) have also been studied in
\cite{gneiting} for the case of positive and bounded functions $V$,
where the decomposition (\ref{vg}) is referred to as \lq\lq scale
mixtures of Euclid's Hat\rq\rq. The improvement of our Theorem
\ref{thm1} compared to \cite{gneiting} lies in its concise proof and
the specific, unified form of the formula for $g$.

The decomposition we derive in the next section is of particular
interest in generalizing results on $N$-body quantum systems so far
only applicable to the Coulomb potential. For instance, it can be used
in estimating the \lq\lq indirect part\rq\rq\ of the interaction
energy, as was done in the Coulomb case
in \cite{LO81}, or to determine the validity of Hartree-Fock
approximations (see \cite{B92} and \cite{GS94}). This is due to the
fact that with the help of (\ref{vg}) the interaction energy in an
$N$-particle quantum system can be decomposed as
\begin{equation}\label{nbody}
\sum_{1\leq i<j\leq N} V(x_i-x_j)= 
\sum_{1\leq i<j\leq N}\int_{\R^n} dz \int_0^\infty dr g(r) 
X^{r,z}_i X^{r,z}_j \ ,
\end{equation}
where $X^{r,z}$ denotes the one-particle multiplication
operator $\chi_r(\,\cdot\,-z)$, being actually a projection, and the
subscript $i$ means that the operator acts on the $i$'th component in
the $N$-fold tensor product appropriate for $N$-particle
systems. Working with the right side of (\ref{nbody}) is often easier
than handling the left side, using the fact that it is a superposition
of products of one-particle projection operators.

In Section \ref{sect3} below we will use the decomposition of the
Yukawa potential to derive the high density asymptotics of the ground
state energy of jellium with Yukawa interaction in the thermodynamic
limit. This is a generalization of \cite{GS94}, where an
analogous expansion in the case of Coulomb interaction was accomplished.

Yukawa potentials are actually used in solid state physics as a model
for the screened Coulomb interaction between electrons in a solid (see,
e.g., \cite{AM}).

\section{The Decomposition}\label{sect2}

In this section, we state and prove the decomposition (\ref{vg}), and
comment on its implications. In the following theorem, we restrict
ourselves to the case $n\geq 2$, and remark on the easy case $n=1$
after the proof.

\begin{thm}[{\bf Decomposition of Radial Functions}]\label{thm1}
For $n\geq 2$, let $V:\R^n\to \R$ be a radial function that is
$[n/2]+2$ times differentiable away from $x=0$. For $m\in\N_0$ denote
$V^{(m)}(|x|)=d^m/d|x|^m V(x)$. Assume that $\lim_{|x|\to\infty}|x|^m
V^{(m)}(|x|)=0$ for all $0\leq m\leq [n/2]+1$, and let
$\chi_r(x)=\theta(r-|x|)$. Then
\begin{equation}\label{vg}
V(x)=\int_0^\infty dr g(r) \chi_{r/2}*\chi_{r/2}(x)\ ,
\end{equation}
where
\begin{eqnarray}\nonumber
g(r)&=&\frac{(-1)^{[n/2]}}{\Gamma(\frac{n-1}2)}\frac 2{(\pi
r^2)^{(n-1)/2}}\\ \nonumber &&\times \Biggl(\int_r^\infty ds 
V^{([n/2]+2)}(s)
\left(\frac d{ds}\right)^{n-1-[n/2]} s (s^2-r^2)^{\half(n-3)}
 \\ \label{defg} &&\qquad +\, \dodd\, V^{([n/2]+2)}(r) r
(2r)^{\half(n-3)} \Gamma(\mbox{$\frac{n-1}2$}) \Biggl)\ ,
\end{eqnarray}
and $\dodd=1$ for $n$ odd, $\dodd=0$ for $n$ even.
\end{thm}

Note the $r/2$ in (\ref{vg}), which is chosen for convenience.

\begin{proof}
Elementary considerations show that
\begin{equation}
\chi_{r/2}*\chi_{r/2}(x)=\frac 1{\Gamma(\frac{n+1}2)}\left(\frac
\pi 4\right)^{(n-1)/2} \int_{|x|}^r dy (r^2-y^2)^{\half(n-1)}
\end{equation}
for $|x|\leq r$, and $0$ otherwise. Inserting the definition
(\ref{defg}) for $g$, we can therefore write
\begin{eqnarray}\nonumber
&&\left( \frac {
(-1)^{[n/2]}}{\Gamma(\frac{n-1}2)\Gamma(\frac{n+1}2)}\frac
1{2^{n-2}} \right)^{-1}\int_0^\infty dr g(r)
\chi_{r/2}*\chi_{r/2}(x)=\\ \nonumber && \int_{|x|}^\infty dr
\frac 1{r^{n-1}} \int_r^\infty ds V^{([n/2]+2)}(s)\\ \nonumber 
&&\times \left(\frac
d{ds}\right)^{n-1-[n/2]} s(s^2-r^2)^{\half(n-3)} \int_{|x|}^r dy
(r^2-y^2)^{\half(n-1)}\\ \nonumber &&+\,\dodd \int_{|x|}^\infty dr
\frac 1{r^{n-1}} V^{([n/2]+2)}(r) r (2r)^{\half(n-3)}
\Gamma(\mbox{$\frac{n-1}2$})\\&&\quad\times\int_{|x|}^r dy 
(r^2-y^2)^{\half(n-1)}\ . 
\end{eqnarray}
We now use the fact that
\begin{eqnarray}\nonumber
&&\left(\frac d{ds}\right)^{n-1-[n/2]} \int_y^s dr \frac
1{r^{n-1}} (r^2-y^2)^{\half(n-1)} s (s^2-r^2)^{\half (n-3)}=\\
\nonumber && \int_y^s dr \frac 1{r^{n-1}} (r^2-y^2)^{\half(n-1)}
\left(\frac d{ds}\right)^{n-1-[n/2]} s (s^2-r^2)^{\half (n-3)}\\
&&+\, \dodd\, \frac 1{s^{n-1}}(s^2-y^2)^{\half(n-1)}s
(2s)^{\half(n-3)}\Gamma(\mbox{$\frac{n-1}2$})\ ,
\end{eqnarray}
and change the order of integration to get
\begin{eqnarray}\nonumber
&&\left( \frac {
(-1)^{[n/2]}}{\Gamma(\frac{n-1}2)\Gamma(\frac{n+1}2)}\frac
1{2^{n-2}} \right)^{-1}\int_0^\infty dr g(r)
\chi_{r/2}*\chi_{r/2}(x)=\\  && \int_{|x|}^\infty ds
V^{([n/2]+2)}(s) \int_{|x|}^s dy \\ \nonumber &&\times \left(\frac
d{ds}\right)^{n-1-[n/2]} \int_y^s dr \frac 1{r^{n-1}}
s(s^2-r^2)^{\half(n-3)}
 (r^2-y^2)^{\half(n-1)}\ .
\end{eqnarray}
The last integral can be evaluated to be
\begin{multline}
\int_y^s dr \frac 1{r^{n-1}} (s^2-r^2)^{\half(n-3)}
(r^2-y^2)^{\half(n-1)}\\ =\frac{\Gamma(\mbox{$\frac{n-1}2$})
\Gamma(\mbox{$\frac{n+1}2$})2^{n-2}(s-y)^{n-1}}{s\Gamma(n)}\ ,
\end{multline}
and therefore
\begin{multline}
\int_0^\infty dr g(r) \chi_{r/2}*\chi_{r/2}(x)\\   =\frac
{(-1)^{[n/2]}}{\Gamma([n/2]+2)}\int_{|x|}^\infty ds
V^{([n/2]+2)}(s) (s-|x|)^{[n/2]+1}= V(x)\ ,
\end{multline}
where we integrated by parts in the last step, using the demanded
decrease properties of $|x|^m V^{(m)}(|x|)$.
\end{proof}

For $n=1$, (\ref{vg}) holds with $g(r)=V^{''}(r)$. This is a well
known result due to P{\'o}lya \cite{polya}, and was used in
\cite{HS01} to get an estimate on the indirect interaction energy in
one-dimensional quantum system.

\begin{rem} If $V(x)$ is $n+1$ times differentiable away from
$x=0$, we can use partial integration to write $g(r)$ in the more
compact form
\begin{equation}\label{comp}
g(r)=\frac{(-1)^{n+1}}{\Gamma(\frac{n-1}2)}\frac 2{(\pi
r^2)^{(n-1)/2}}  \int_r^\infty ds V^{(n+1)}(s) s
(s^2-r^2)^{(n-3)/2} \ .
\end{equation}
\end{rem}

For simplicity, we have restricted ourselves to formulating Theorem
\ref{thm1} for differentiable functions only, but as is obvious from
the proof, (\ref{defg}) and (\ref{comp}) hold for a more general class
of potentials, if the derivatives are interpreted in the sense of
distributions.

\begin{ex} 
A particular simple example which our decomposition applies to is the
Coulomb potential $V(x)=1/|x|$. For the general case of $n$ dimensions
we compute the weight function to be
\begin{equation}
g(r)=\frac 12 \frac{\Gamma(n+2)}{\Gamma(n/2+1)}
\frac{\pi^{1-n/2}}{r^{n+2}}\ .
\end{equation}
\end{ex}

\begin{ex} \label{ex2}
Another example, which will be used below, is the Yukawa potential
$Y_\mu(\x)=\exp(-\mu|\x|)/|\x|$, with $\mu>0$. In three dimensions, the
corresponding weight function reads
\begin{equation}
g(r)=\frac 2\pi \frac {e^{-\mu r}}{r^5} \left[ 8 + 8\mu r + 4 (\mu
r)^2 + (\mu r)^3\right] \ .
\end{equation}
\end{ex}

The decomposition (\ref{vg}) provides conditions for positive
definiteness of $V$. By positive definiteness of a locally integrable
function $V$ we mean that
\begin{equation}
\int_{\R^n\times\R^n}  V(x-y) {\overline {\varphi(x)}} \varphi(y)dxdy \geq 0
\end{equation}
for all $\varphi\in C_0^\infty(\R^n)$. This is equivalent to $V$
having a positive Fourier transform in the distributional sense.
Since $\chi_r*\chi_r(x)$ is obviously positive definite (as a function
of $x$), any $V$ with a positive weight function $g(r)$ is positive
definite. It is easy to see that
\begin{equation}\label{pos}
\left(\frac d{ds}\right)^{n-1-[n/2]} s (s^2-r^2)^{\half(n-3)}\geq
0 \quad \mbox{ for }s\geq r\ ,
\end{equation}
and therefore positivity of $(-1)^{[n/2]} V^{([n/2]+2)}$ implies that
$V(x)$ is positive definite. This is a result due to Askey
\cite{A73a}, who deduces it from certain positivity properties of
Bessel functions proven in \cite{A73b} and \cite{FI75}. Further
extensions and refinements can be found in \cite{gneiting2} and
references therein.

Consider now the (physically interesting) special case $n=3$. Here the
weight function $g$ takes the simple form
\begin{equation}
g(r)=\frac 2{\pi r^2}\left(V^{''}(r)-r V^{'''}(r)\right)=-\frac 2\pi
\left(\frac {V^{''}(r)}{r}\right)^{'}\ .
\end{equation}
Hence a monotone decrease of $V^{''}/r$ (instead of $V^{''}$) is
sufficient for positive definiteness of $V$. This condition for
positive definiteness of a radial function on $\R^3$ has been obtained
before by A. Martin \cite{GM97}. See also \cite{HN} for an earlier
reference.

Positivity of $g$ might be a useful way of checking positive
definiteness of a function $V$. However, this condition is of
course not necessary, as the following example shows.

\begin{ex}
For $n=3$, $V(x)=1/(|x|^2+1)$ is positive definite, but the
corresponding $g(r)$ is not a positive function.
\end{ex}

\section{Ground State Energy of Yukawa Jellium}\label{sect3}

We now demonstrate the usefulness of our decomposition (\ref{vg}) in
applications to $N$-body quantum systems. We consider a model of
electrons interacting with Yukawa potentials and with a uniform
charged background. Yukawa potentials are used in solid state physics
as a model for screened Coulomb potentials \cite{AM}. They were first
introduced by Debye and H\"uckel \cite{DH}, and were later used in
meson theory by Yukawa.

The system under consideration is described by the Hamiltonian
\begin{equation}
H=\sum_{i=1}^N \left( -\Delta_i + V(\x_i) \right) + \sum_{1\leq
i<j\leq N} Y_\mu(\x_i-\x_j) \ ,
\end{equation}
acting on $\Hh=\bigwedge_{i=1}^N L^2(\Lambda,d\x_i;\C^q)$, the antisymmetric tensor product of the one-particle space $L^2(\Lambda,d\x_i;\C^q)$ of one electron with spin $(q-1)/2$. The
particles are confined to the cube $\Lambda=[0,L]^3$ of side length
$L$, with volume $|\Lambda|=L^3$. 
They interact via the Yukawa potential
\begin{equation}
Y_\mu(\x)=\frac{e^{-\mu |\x|}}{|\x|} \ ,
\end{equation}
with $\mu>0$. The electrons move in the potential $V$ created by
a charged background, given by
\begin{equation}
V(\x)=\rho \int_\Lambda d\y \, Y_\mu(\x-\y) \ ,
\end{equation}
where $\rho>0$ is the background density. We assume neutrality of the system, i.e., $\rho=N/|\Lambda|$.  

We have to specify boundary conditions for $H$ on $\Lambda$, and we
take Dirichlet boundary conditions for simplicity. Denoting
$E(N,\Lambda,\mu)={\rm inf spec\, } H$, we define the {\it energy
density} for a neutral system in the thermodynamic limit as
\begin{equation}\label{defe}
e(\rho,\mu)=\lim_{L\to\infty} \frac{1}{|\Lambda|} \left( E(\rho
|\Lambda|, \Lambda, \mu) +
\half \rho^2 \int_{\Lambda\times\Lambda} d\x d\y \, Y_\mu(\x-\y) \right) \ .
\end{equation} 
For convenience, we have added the self energy of the background
charge. For $\mu=0$ this is essential, because otherwise the limit does not exist. We take the existence of the limit (\ref{defe}) for granted. For the
Coulomb case (i.e. $\mu=0$), existence (and independence of the
boundary conditions for $H$) of the thermodynamic limit has been
shown in \cite{LN75}. However, because of the short-range nature of the Yukawa
potential, the existence of the thermodynamic limit for $\mu>0$ is much easier
to show.

Our main theorem concerns the behavior of $e(\rho,\mu)$ for large
$\rho$. This is an extension of the result in \cite{GS94} for $\mu=0$.
We denote by $\kF$ the {\it Fermi momentum}
\begin{equation} 
\kF=(6\pi^2\rho/q)^{1/3} \ .
\end{equation}

\begin{thm}[{\bf Ground State Energy of Yukawa Jellium}]\label{jelli}
As $\rho\to\infty$,
\begin{equation}\label{asy}
\frac {e(\rho,\mu)}\rho=\frac 35 \kF^{2} -
\frac 3{4\pi}\kF 
J\left(\frac \mu\kF \right)\big[1 +
o(1)\big] \ ,
\end{equation}
uniformly in $\mu$ for bounded $\mu/\kF$. Here $J$ is
the function
\begin{equation}
J(\eta)=1-\frac 16 \eta^2 + \frac12\left(\eta^2+\frac 1{12}
\eta^4\right)\ln\left(1+\frac 4{\eta^2}\right) -
\frac 43 \eta\arctan\frac 2\eta \ .
\end{equation}
\end{thm}

As in \cite{GS94}, o(1) is positive and smaller than
$O(\rho^{-1/15+\eps})$ for any $\eps>0$ and for $\mu/\kF$ fixed. The
second term on the right side of (\ref{asy}) is entirely due to
correlations in the ground state wave function and is usually referred to as
\lq\lq exchange energy\rq\rq.
The function $J(\eta)$ in Theorem~\ref{jelli} is in fact the integral
\begin{equation}
J(\eta)=\frac 43 \int_0^2 dk \frac {k^2}{k^2+\eta^2} \left(1-\frac 34
k+\frac 1{16} k^3\right)\ .
\end{equation}
Notice that $J(0)=1$, whereas $J(\eta)\approx
4/(9\eta^2)$ as $\eta\to\infty$.

We conjecture that the asymptotics in (\ref{asy}) is uniform in
$\mu$, even if $\mu/\kF$ tends to infinity. For $q=1$, i.e., the
spinless case, this is in fact easy to prove, as will be remarked after
the proof of Theorem~\ref{jelli}.
Note, however, that in physical situation $\mu/\kF$ is small. The range of the potential is given by $1/\mu$, which is always bigger than the mean particle distance $\rho^{-1/3}$. 

The proof of Theorem~\ref{jelli} follows essentially the analogous discussion of the Coulomb
case in \cite{GS94}. We use freely the estimates derived there, mainly
pointing out the differences to our case.

\begin{proof}
To obtain an upper bound to the ground state energy, we use as a
trial state a Slater determinant of the $N$ lowest eigenvectors of the
Laplacian on $\Lambda$, including spin. The calculation of the
expectation value is essentially the same as the one done by Dirac
\cite{dirac} in the Coulomb case. In the thermodynamic limit the
boundary conditions do not matter (see \cite{GS94} for details), and
one can just consider plain waves, with momenta up to the Fermi
momentum $\kF$. The function $J$ then arises from the integral
\begin{equation}
\frac q2 \int_{\R^3} \frac {dk}{(2\pi)^3} \widehat Y_\mu (k) 
\int_{\R^3} \frac {dq}{(2\pi)^3} \theta(\kF-|q|) \theta(\kF-|k-q|) = 
\frac 3{4\pi}\rho\, \kF 
J\left(\frac \mu\kF \right) \ .
\end{equation}

We are left with the lower bound. We start with the following
decomposition of the Yukawa potential, proven in Section
\ref{sect2}. If $\chi_r$ denotes the characteristic function of a ball
of radius $r$ centered at the origin, then
\begin{equation}\label{decomp}
Y_\mu(\x-y)=\int_{\R^3} d\z \int_0^\infty dr \widetilde g(r) \chi_r(\x-\z) 
\chi_r(\y-\z) \ ,
\end{equation}
where
\begin{equation}
\widetilde g(r)= 2 g(2r) =\frac 1\pi \frac {e^{-2\mu r}}{r^5} \left[ 1 + 2\mu r + 2 (\mu
r)^2 + (\mu r)^3\right] 
\end{equation}
(compare with Example~\ref{ex2}). 
The key to the lower bound is a correlation inequality derived in
\cite[Cor.~5]{GS94}.  It states that for any projection $X$ and $0\leq P\leq
1$ on the one-particle space, and for antisymmetric $\psi\in\Hh$,
\begin{eqnarray}\nonumber
&&\left\langle \psi \left| \sum_{1\leq i<j\leq N} X_i X_j
\right|\psi\right\rangle \geq
\half \Tr[X\gamma]^2-\half \Tr[PXPX]\\ &&-\const 
\Tr[X(P+\gamma)]\min\left\{1,\left(\Tr[X(1-P)\gamma(1-P)]\right)^{1/2} 
\right\} \ .
\end{eqnarray}
Here we denote by $\gamma$ the one-particle reduced density matrix
of $\psi$, with corresponding density $\rho_\gamma$. Using this, with $X=\chi_r(\,\cdot\, - \z)$, and integrating
over $r$ and $\z$ as in (\ref{decomp}), we get that
\begin{eqnarray}\nonumber
&&\left\langle \psi \left| \sum_{1\leq i<j\leq N} Y_\mu(\x_i-\x_j)
\right|\psi\right\rangle
\geq 
\half \int_{\Lambda\times\Lambda} d\x d\y \rho_\gamma(\x)
\rho_\gamma(y) Y_\mu(x-y) \\ \label{28} &&- 
\half \sum_{\sigma,\sigma'}\int_{\Lambda\times\Lambda} d\x d\y 
|P(\x,\sigma;y,\sigma')|^2
Y_\mu(x-y)- {\rm Error} \ .
\end{eqnarray}
The function $P(\x,\sigma;y,\sigma')$ is the integral kernel of $P$, and $\sigma \in (1,\dots,q)$ denotes the spin variables of the electrons. The error term is
\begin{multline}
{\rm Error} = \\ \const \int_{\R^3} d\z \int_0^\infty dr \widetilde g(r)
\big[\chi_r*(\rho_\gamma+\rho_P)\big](\z)
\min\left\{1,\big[\chi_r*\rho_{Q\gamma Q}\big](\z)^{1/2}\right\} \ .
\end{multline}
Here we introduced the operator $Q=1-P$; $\rho_P$ and $\rho_{Q\gamma Q}$ are the densities corresponding to $P$ and $Q\gamma Q$, respectively.

Since $\widetilde g(r)\leq {\rm
const.\, } r^{-5}$, we can proceed exactly as in \cite[Lem.~6]{GS94} to
estimate
\begin{equation}\label{30}
{\rm Error}\leq {\rm const.\, } \|\rho_\gamma+\rho_P\|_1^{1/6+\eps} \|\rho_\gamma+\rho_P\|_{5/3}^{5/6}\, \delta(\gamma,P)^{1/3-\eps}
\end{equation} 
for any $0<\eps\leq 1/6$. Here $\delta(\gamma,P)=\Tr[\gamma(1-P)]$ measures the \lq\lq difference\rq\rq\ of $\gamma$ and $P$. Since $\psi$ is supposed to be the ground state, we can use the upper bound to get the {\it a priori} knowledge that $\|\rho_\gamma\|_{5/3}^{5/3}\leq \const N \kF^2$, see \cite[Eq.~(4.11)]{GS94}.  

Now let $P$ be the projection onto the first $N$ eigenstates
of the Laplacian with periodic boundary conditions. It is shown in \cite[Eq.~(4.13)]{GS94} that, for any $\psi\in\Hh$,
\begin{equation}\label{31}
\frac 1N \left\langle\psi \left| -\sum_{i=1}^N \Delta_i \right|\psi\right\rangle \geq  \kF^2\left(\frac 35 + \const  \big(N^{-1} \delta(\gamma,P) \big)^2 \right)- o(1)  
\end{equation}
as $|\Lambda|\to \infty$. The second term in (\ref{28}), when
divided by $N$, converges in the thermodynamic limit to
$\mbox{$\frac3{4\pi}$} \kF J(\mu/\kF)$, as explained in the upper
bound. Moreover, $\|\rho_P\|_{5/3}^{5/3}\leq \const N \kF^2$.  Putting together (\ref{28}), (\ref{30}) and (\ref{31}), we therefore get that in the ground
state $\psi$,
\begin{multline}\label{des}
\frac 1N  \left\langle\psi \left| H \right|\psi\right\rangle + \frac 1{2N} 
 \rho^2 \int_{\Lambda\times\Lambda} d\x d\y \, Y_\mu(\x-\y) 
\\ \geq \kF^2\left(\frac 35 + \const  \big(N^{-1} \delta(\gamma,P) \big)^2 \right) - \frac3{4\pi} \kF J(\mu/\kF) \\- \const \kF  \big(N^{-1} \delta(\gamma,P) \big)^{1/3-\eps} \\ +\half \int_{\Lambda\times\Lambda} d\x d\y \big(\rho_\gamma(\x)-\rho\big)\big(
\rho_\gamma(y)-\rho\big) Y_\mu(x-y) - o(1) 
\end{multline}
as $|\Lambda|\to\infty$. 
The last term is positive, since $Y_\mu$ is positive definite. Minimizing the right side of (\ref{des}) over $\delta(\gamma,P)$ gives the desired result. 
\end{proof}

\begin{rem} 
We expect the asymptotics in (\ref{asy}) to be uniform in $\mu/\kF$
even if this value goes to infinity. The reason is that for
$\mu/\kF\gg 1$ the interaction potential is of very short range, and hence 
the interaction between particles of the
same spin can be neglected to leading order due to the antisymmetry of the wave
function. I.e., in this case one expects a contribution to
$e(\rho,\mu)$ from the interaction energy of $(1-q^{-1}) 2\pi
\rho^2/\mu^2$. The total energy for $\mu\gg \rho^{1/3}$ should then be
\begin{equation}
\frac {e(\rho,\mu)}{\rho}\approx \frac 35 \kF^2 - \frac {4\pi}{\mu^2}\rho + \big(1-q^{-1}\big) \frac {2\pi}{\mu^2}\rho + \frac {2\pi}{\mu^2}\rho \ ,
\end{equation} 
which is exactly the same as (\ref{asy}) as $\mu/\kF\to \infty$. 

For the spinless case, i.e., $q=1$, the  uniformity is in fact easy to prove. Neglecting the positive interaction energy we get a lower bound
\begin{equation}
\frac {e(\rho,\mu)}{\rho}\geq \frac 35 \kF^2 - \frac {2\pi}{\mu^2}\rho \ ,
\end{equation} 
which agrees with (\ref{asy}) as $\mu/\kF\to \infty$. We note that the
second term is now not the exchange energy, but the difference
of the energy of the electrons in the background and the self energy
of the background.
\end{rem}

\bigskip
\noindent {\it Acknowledgments.} 
We thank Volker Bach for encouraging the present study, and Michael
Loss for providing the references \cite{A73a, A73b, FI75}. C.H.  was
supported by a Marie Curie Fellowship of the European Community
programme \lq\lq Improving Human Research Potential and the
Socio-economic Knowledge Base\rq\rq\ under contract number
HPMFCT-2000-00660. R.S. was supported by the Austrian Science Fund,
and acknowledges warm hospitality at the Mathematical Institute, LMU
M\"unchen, where part of this work was done.

\end{document}